\documentclass[fleqn,usenatbib,useAMS]{mnras}

\usepackage{newtxtext,newtxmath}
\usepackage[T1]{fontenc}

\newcommand{\chieff}{\chi_\text{eff}}

\usepackage{graphicx}
\usepackage{amsmath}
\usepackage{color}

\title[Do unequal-mass BBH systems have larger $\chi_\text{eff}$?]{
Do unequal-mass binary black hole systems have larger $\chi_\text{eff}$? \\
Probing correlations with copulas in gravitational-wave astronomy}

\author[Adamcewicz \& Thrane]{
Christian Adamcewicz$^{1,2}$\thanks{E-mail: cada0007@student.monash.edu}
and Eric Thrane$^{1,2}$
\\
$^{1}$School of Physics and Astronomy, Monash University, Clayton VIC 3800, Australia\\
$^{2}$OzGrav: The ARC Centre of Excellence for Gravitational Wave Discovery, Clayton VIC 3800, Australia
}

\date{Accepted XXX. Received YYY; in original form ZZZ}
\pubyear{2022}

\begin{document}
\label{firstpage}
\pagerange{\pageref{firstpage}--\pageref{lastpage}}
\maketitle

\begin{abstract}
The formation history of binary black hole systems is imprinted on the distribution of their masses, spins, and eccentricity.
While much has been learned studying these parameters \textit{in turn}, recent studies have explored the \textit{joint} distribution of binary black hole parameters in two or more dimensions.
Most notably, it has recently been argued that binary black hole mass ratio and effective inspiral spin $\chi_\text{eff}$ are anti-correlated.
We point out a previously overlooked subtlety in such two-dimensional population studies: in order to conduct a controlled test for correlation, one ought to fix the two marginal distributions---lest the purported correlation be driven by improved fit in just one dimension.
We address this subtlety using a tool from applied statistics: the copula density function.
We use the previous work correlating mass ratio and $\chi_\text{eff}$ as a case study to demonstrate the power of copulas in gravitational-wave astronomy while scrutinising their astrophysical inferences.
Our findings, however, affirm their conclusions that binary black holes with unequal component masses exhibit larger $\chi_\text{eff}$ (98.7\% credibility).
We conclude by discussing potential astrophysical implications of these findings as well as prospects for future studies using copulas.
\end{abstract}


\section{Introduction}
\label{sec:introduction}
There is still much uncertainty surrounding the astrophysics of binary black hole (BBH) formation \citep[see the reviews by][for example]{Mapelli_2018, Mandel_2022, Spera_2022}.
Nonetheless, some general features can be associated with different formation channels.
Binaries assembled in the field are likely to contain black holes with spin vectors that are preferentially aligned to the orbital angular momentum vector \citep{Mandel_2016, Belczynski_2016b, Stevenson_2017, Marchant_2016, Talbot_2017, Qin_2018}.
The vast majority of these systems should be nearly circular (negligible eccentricity) \citep{Peters_1964, Hinder_2008} and they ought to include black holes with masses below the pair instability gap \citep{Belczynski_2016, Marchant_2019, Stevenson_2019, Woosley_2021}.
Binaries assembled dynamically in dense stellar environments are likely to contain black holes with isotropic random spin orientations \citep{Schnittman_2004, Bogdanovic_2007, Rodriguez_2015, Rodriguez_2016b, Stone_2016, Yu_2020}.
Some fraction of these systems, perhaps $\approx 5\%$, may be measurably eccentric \citep{Samsing_2014, Samsing_2017, Samsing_2018, Lower_2018, Zevin_2019, Romero-Shaw_2019, Gondan_2021}.
These binaries may include `second-generation' black holes in the pair instability mass gap \citep{Oleary_2016, Fishbach_2017, Gerosa_2017, Rodriguez_2018, hierarchical, gwtc2_hierarchical, Doctor_2020}.
By fitting the observed distribution of BBH masses, spins, and eccentricity, it is possible to estimate the fraction of BBH systems assembled through each of these channels \citep[see][and references therein]{GWTC1_analysis, GWTC2_analysis, GWTC3_analysis, Kaze_2021, Farr_2018, Baibhav_2020, Zevin_2021, Bouffanais_2021, Roulet_2021}.

To date, the majority of population modelling in gravitational-wave astronomy has been carried out using models where parameters factorise into statistically independent distributions.
This was a reasonable starting point, providing a simple framework for the analysis of a relatively small dataset.
However, as the number of gravitational-wave detections has increased, there has been growing interest in multivariate models, which allow for two or more parameters to be correlated; namely effective inspiral spin $\chieff$ with mass \citep{Safarzadeh_2020, Hoy_2022, Biscoveanu_2022, Franciolini_2022, Tiwari_2022}, $\chieff$ with mass ratio \citep{Callister_2021, Franciolini_2022, Tiwari_2022}, $\chieff$ with redshift \citep{Biscoveanu_2022, Bavera_2022, Tiwari_2022}, and redshift with mass \citep{Fishbach_2021, Belczynski_2022}.
This shift is motivated by theoretical studies, predicting multivariate distributions in $\chieff$ and mass (as well as mass ratio) \citep{Bavera_2021, Bavera_2022, Broekgaarden_2022}, $\chieff$ and redshift \citep{Bavera_2020}, redshift and mass \citep{Neijssel_2019, van_Son_2022, Mapelli_2022}, as well as mass ratio and component spin \citep{Lousto_2008}.
In \cite{Kruckow_2021}, the authors also propose that there is a correlation between the total mass and chirp mass.
However, such a correlation would seem hard to avoid barring a pathological relationship between mass ratio and total mass.

We highlight one notable study, being \cite{Callister_2021}, which finds evidence for an anti-correlation between mass ratio and $\chieff$.
This was somewhat surprising, as theoretical studies prior to this work had only predicted such a multivariate distribution in the case of two rather specific BBH formation channels.
The first of these is formation via common envelope evolution in binaries with very high common envelope efficiency, whilst the second is via stable mass transfer when supposing super-Eddington accretion \citep{Bavera_2021, Zevin_2022}.
Follow-up studies suggested that this anti-correlation may also occur in BBH systems that undergo mass ratio reversal during stable mass transfer \citep{Broekgaarden_2022}, as well as in BBH systems assembled dynamically in active galactic nuclei (AGN) \citep{Mckernan_2022}.
If it is real, this covariance is an important clue for our understanding of BBH formation.

In this paper, we point out an important subtlety associated with multivariate population models.
We are particularly concerned with studies that seek to establish a correlation $\kappa$ between two parameters $(x,y)$.
The details are somewhat technical (see Section~\ref{sec:subtleties_in_covariant_models}), but the basic idea is that models in previous studies are constructed so that our assumptions about the distribution of $y$ depend on the correlation parameter $\kappa$.
This is not ideal because it means two knobs are always being changed at the same time: the degree of correlation and the shape of the $y$ distribution.
As a consequence, if one value of $\kappa$ is preferred over another, it is not clear if this is due to a bona fide correlation or due to an improved fit to the shape of the $y$ distribution.
We show how this problem is solved using a tool from applied statistics: copulas \citep[detailed in][]{Sklar_1996}.

Copulas have seen extensive application modelling covariance, especially in finance \citep[see][for an overview]{Bouye_2000}.
Here, we present a few examples of copula applications in astronomy.
First, copulas have been used to model covariance between bands in luminosity functions \citep{Takeuchi_2010, Takeuchi_2013, Andreani_2014, Yuan_2018}, model multivariate mass functions \citep{Takeuchi_2020}, and model mass-luminosity correlations \citep{Gunawardhana_2014, Andreani_2018} in galaxies.
Copulas have also been used in cosmology---estimating cosmological parameters from gravitational lensing surveys \citep{Sato_2010, Lin_2015}, and modelling the dark matter density field \citep{Scherrer_2009, Qin_2020}.
As a final example, copula models have been used to look for period-mass relations in extrasolar planetary systems \citep{Jiang_2008}.

The remainder of this paper is organized as follows.
In Section~\ref{sec:subtleties_in_covariant_models}, we investigate the challenges of two-dimensional models, using \cite{Callister_2021} as a case study.
In Section~\ref{sec:an_introduction_to_copulas}, we show how these challenges are addressed with copulas.
In Section~\ref{sec:model_and_implementation}, we use this framework to propose a new model for the distribution of BBH mass ratio and effective inspiral spin.
We present our results in Section~\ref{sec:results}.
We conclude in Section~\ref{sec:discussion} and discuss how this work may be extended in the future.

\section{The challenges of two-dimensional models}
\label{sec:subtleties_in_covariant_models}

\begin{figure*}
    \centering
    \includegraphics[width=\textwidth]{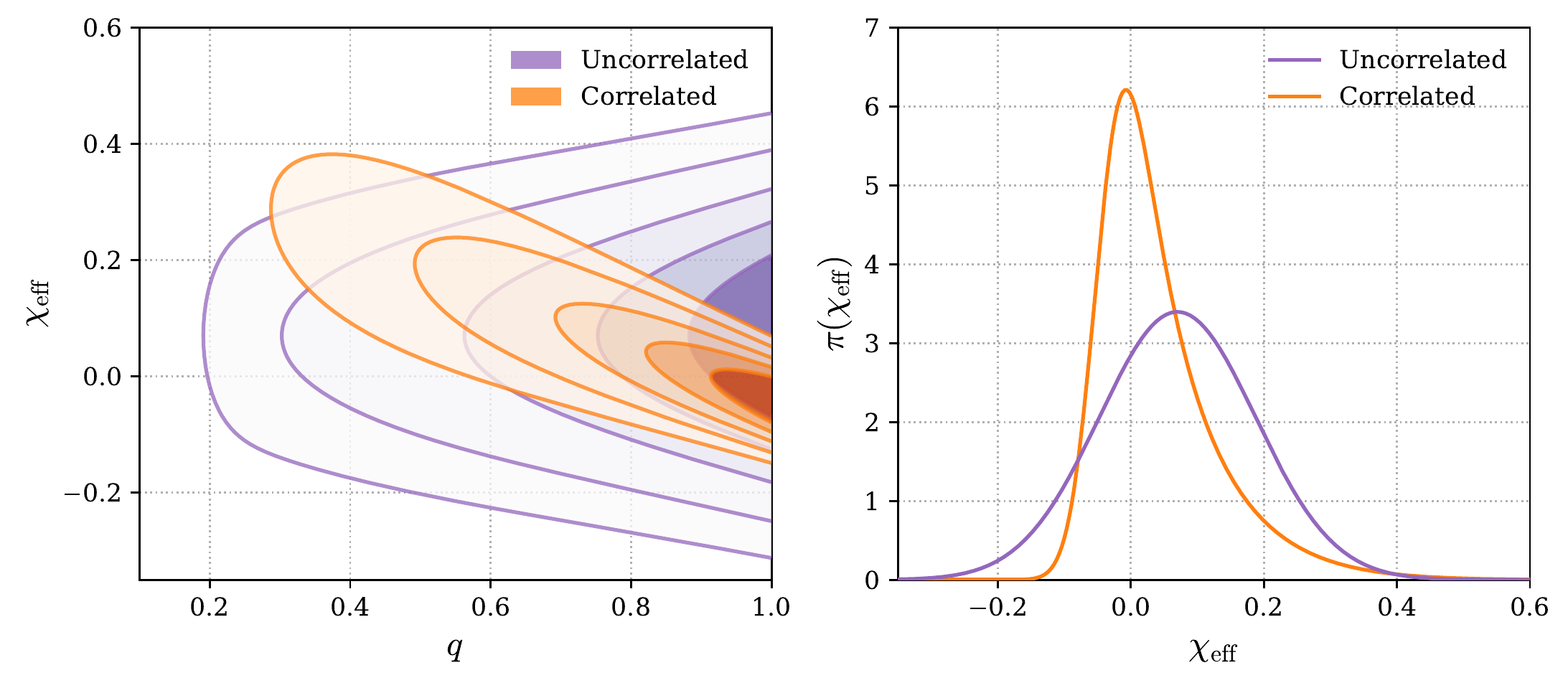}
    \caption{Reconstructed distributions using the best-fit parameters from Callister \textit{et al.} (2021). In purple is the uncorrelated model, whilst the correlated model is shown in orange. \textbf{Left}: the joint distributions for mass ratio and effective inspiral spin. \textbf{Right}: the marginal distributions for effective inspiral spin. Note how the act of marginalisation produces a highly skewed, non-Gaussian distribution for the orange correlated model.}
    \label{fig:callister+}
\end{figure*}

We are interested in the class of problems where one seeks to establish if there is a correlation present in two parameters $(x,y)$.
As a case study, we dissect the model from \cite{Callister_2021}, which is focused on the parameters of mass ratio
\begin{equation}
    q \equiv \frac{m_2}{m_1},
\end{equation}
and effective inspiral spin \citep{Damour_2001}
\begin{equation} \label{eq:chi_eff}
    \chieff = \frac{\chi_1 \cos t_1 + q \chi_2 \cos t_2}{1 + q} ,
\end{equation}
though, the problem is a quite general one.
Here, $m_1$ and $m_2$ are the primary and secondary component masses of the binary.
The mass ratio is allowed to vary on the interval $[m_{\min}/m_1, 1]$.
Meanwhile, $\chi_1$ and $\chi_2$ are the corresponding dimensionless spin magnitudes, and $t_1$ and $t_2$ are the tilt angles between each spin vector and the orbital angular momentum.

The first step in \cite{Callister_2021} \citep[and similar gravitational-wave studies;][]{Safarzadeh_2020, Biscoveanu_2022} is to make some assumption about the distribution of one parameter, in this case, mass ratio. 
\cite{Callister_2021} adopt a power-law distribution
\begin{equation}
\label{eq:p_q_m1}
    \pi(q|m_1,\Lambda) \propto q^\gamma .
\end{equation}
Here, $\gamma$ is one of the several hyper-parameters describing the BBH population---the set of which we denote $\Lambda$.
Specifically, $\gamma$ controls the shape of the marginal distribution for mass ratio.
As the distribution of mass ratio is subtly conditioned on primary mass, it is worth noting that the primary mass distribution follows the power-law and peak model proposed by \cite{Talbot_2018}, such that
\begin{equation}
\label{eq:p_m1}
    \pi(m_1|\Lambda) = f_p \mathcal{P}(m_1|\Lambda) + (1 - f_p) \mathcal{N}(m_1|\Lambda) .
\end{equation}
Here,
\begin{equation}
    \mathcal{P}(m_1|\Lambda) \propto m_1^\lambda,
\end{equation}
is a power-law,
\begin{equation}
    \mathcal{N}(m_1|\Lambda) \propto \exp \left[-\frac{(m_1 - \mu_m)^2}{2 \sigma_m^2}\right],
\end{equation}
is a Gaussian feature, and $f_p$ is a hyper-parameter that denotes the fraction of BH masses in the power-law component.
Of course, $\lambda$, $\mu_m$, and $\sigma_m$ are also hyper-parameters.

The next step is to assume some functional form for the distribution of the second variable \textit{conditioned on the first variable}.
\cite{Callister_2021} assume
\begin{equation}
\label{eq:gaussian_p_chi_eff}
    \pi(\chieff|q, \Lambda) \propto \exp\left[-\frac{\Big(\chieff - \mu_\chi(q, \Lambda)\Big)^2}{2 \sigma_\chi(q,\Lambda)^2}\right].
\end{equation}
This is a Gaussian distribution with a mean that depends on the mass ratio:
\begin{equation}
\label{eq:correlated_mu_chi}
    \mu_{\chi}(q,\Lambda) \equiv \mu_{\chi,0} + \alpha (q - 0.5) .
\end{equation}
Here, $\mu_{\chi, 0}$ and $\alpha$ are two more hyper-parameters.
The $\alpha$ parameter controls the degree of covariance; the covariance vanishes when $\alpha=0$.
The width of the Gaussian is also conditioned on mass ratio:
\begin{equation}
\label{eq:correlated_sigma_chi}
     \log_{10}\sigma_{\chi}(q,\Lambda) \equiv \log_{10}\sigma_{\chi,0} + \beta (q - 0.5).
\end{equation}
Here, $\sigma_{\chi,0}$ and $\beta$ are also hyper-parameters.

The model is fit to LIGO--Virgo \citep{LIGO, Virgo} data, using GWTC-2 \citep{GWTC2, data} in order to obtain posterior distributions for the hyper-parameters describing the distribution of BBH mass and spin.
\cite{Callister_2021} finds support for a negative value of $\alpha = -0.46_{-0.28}^{+0.29}$ (90\% credibility).
These results seem to imply a significant anti-correlation between mass ratio and effective inspiral spin, as seen in the left panel of Fig.~\ref{fig:callister+}.
In fact, a value of $\alpha \geq 0$ is ruled out by \cite{Callister_2021} with a credibility of 98.7\%.

However, given the way this model is constructed, different hyper-parameter values for $\alpha$ or $\beta$ produce different marginal distributions for $\chi_\text{eff}$:
\begin{equation}
\label{eq:p_chi_eff}
    \pi(\chieff|\Lambda) = \int_{q_{\min}}^1 dq \, \pi(\chieff|q,\Lambda) \pi(q|\Lambda) .
\end{equation}
This is dramatically illustrated  in Fig.~\ref{fig:callister+}, which shows two different marginal distributions $\pi(\chieff|\Lambda)$---one with an $\alpha<0$ anti-correlation (orange) and one with no correlation (purple).
Introducing covariance between $(q, \chieff)$ has reshaped the marginal distribution of $\chieff$.
This begs the question: \textit{why} do the data prefer $\alpha<0$---is it because there is truly a correlation between $(q, \chieff)$; or is it because the $\chieff$ distribution in nature is better described as asymmetric (orange) than symmetric (purple); or is it some combination of both factors?

While this is a potential issue with all correlation studies that utilise this framework, the case of $q$ and $\chieff$ is particularly interesting.
As noted in \cite{Callister_2021}, their asymmetric fit for $\pi(\chieff|\Lambda)$ (see Fig.~\ref{fig:callister+}) is qualitatively similar to the results of \cite{Roulet_2021}.
Furthermore, \cite{Galaudage_2021b} finds a significant zero-spin peak in the distribution of BBH spin magnitudes ($\chi_1$ and $\chi_2$), although follow-up studies in \cite{Callister_2022} and \cite{Mould_2022} find no such feature.
Regardless, this may raise the question of whether the orange correlated marginal distribution in Fig.~\ref{fig:callister+} is favourable, due to increased support near $\chieff = 0$, when compared to the purple uncorrelated case.
We therefore seek to determine if the correlation observed by \cite{Callister_2021} can be explained away as an artifact of their model construction.

\section{Copulas}
\label{sec:an_introduction_to_copulas}

\begin{figure*}
    \centering
    \includegraphics[width=\textwidth]{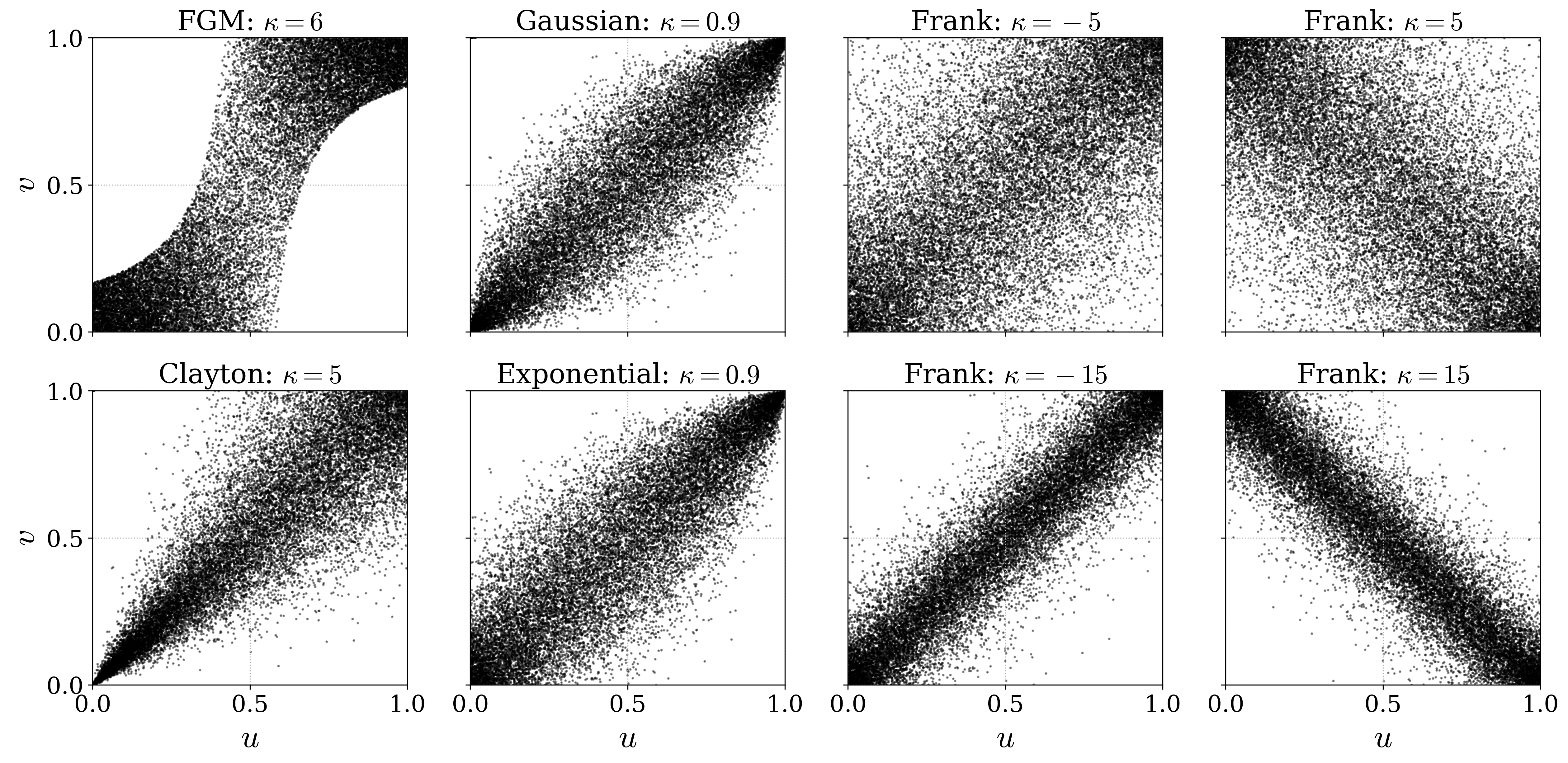}
    \caption{Examples of copula density functions. The different values of $\kappa$ yield different degrees of correlation. Note in every case, the marginal distributions for $u$ and $v$ are uniformly distributed. When $\kappa=0$, there is no correlation present, which would correspond to a two-dimensional uniform distribution (a featureless square). As $\kappa$ increases in magnitude, a correlation or an anti-correlation is introduced, depending on the sign of $\kappa$ and the type of copula density function used. Listed above each example is the name of the copula density function and the value of $\kappa$ used.}
    \label{fig:copula_examples}
\end{figure*}

In order to address the challenges identified in the previous section, we seek to construct a model in which covariance can be introduced without changing the marginal distributions for either $(q, \chieff)$.
Copulas are a tool from applied statistics for exactly this purpose \citep{Sklar_1996}.
In this section, we provide an introduction to copulas.
This will provide the foundation upon which we will build our new model for $(q, \chieff)$ in Section~\ref{sec:model_and_implementation}.
Whilst copulas traditionally refer to cumulative distribution functions \citep[see][]{Sklar_1996}, we use the term as a short-hand for their related copula density functions.
Formally, a copula density function is any joint probability distribution $\pi_c(u,v|\kappa)$, for parameters $u,v \in [0,1]$, with \textit{uniform} marginal distributions for both $u$ and $v$
\begin{align}
    \pi(u|\kappa) = & \int_0^1 dv \, \pi_c(u,v|\kappa) = 1 \\
    \pi(v|\kappa) = & \int_0^1 du \, \pi_c(u,v|\kappa) = 1 ,
\end{align}
conditioned on hyper-parameter $\kappa$, which controls the degree of correlation between $(u,v)$.
Applied statisticians have derived a large variety of copulas, each providing a unique shape for the joint distribution when $\kappa\neq0$.
Different shapes are suitable for different physical models.
Examples of different copulas are shown in Fig.~\ref{fig:copula_examples}.

In general, physical parameters are not distributed uniformly on the interval $[0,1]$.
However, physical parameters $(x,y)$ are easily related to copula parameters $(u,v)$ using cumulative density functions:
\begin{align}
\label{eq:example_u}
    u(x) = \int_{x_{\min}}^x dx' \, \pi(x') \\
\label{eq:example_v}
    v(y) = \int_{y_{\min}}^y dy' \, \pi(y')  .
\end{align}
Thus, we may construct a distribution for $(x,y)$ as
\begin{equation}\label{eq:x-y_to_u-v}
    \pi_c(x,y|\kappa) = \pi(x) \pi(y) \pi_c\big(u(x),v(y)|\kappa\big) ,
\end{equation}
which preserves the (non-uniform) marginal distributions for $x$ and $y$, while allowing the covariance to change according to $\kappa$.
Technically, $\pi_c(x,y|\kappa)$ is not a copula because $x$ and $y$ are not uniformly distributed, but as shorthand, we refer to this distribution as a copula for $(x,y)$.

\section{Model and implementation}
\label{sec:model_and_implementation}

\begin{table*}
    \centering
    \begin{tabular}{c l l}
        \hline
        Parameter & Prior & Description \\
        \hline
        \hline
        $\mu_{\chi,0}$ & $\mathcal{U}(-1, 1)$ & 
        Mean of Gaussian $\chi_\text{eff}$ distribution when $\alpha=0$\\
        \hline
        $\log_{10}\sigma_{\chi,0}$ & $\mathcal{U}(-1.5, 0.5)$ & 
        Log standard deviation of Gaussian $\chi_\text{eff}$ distribution when $\beta=0$\\
        \hline
        $\alpha$ & $\mathcal{U}(-2.5, 1)$ & 
        Reshapes the $\chi_\text{eff}$ distribution through $q$\\
        \hline
        $\beta$ & $\mathcal{U}(-2, 1.5)$ & 
        Reshapes the $\chi_\text{eff}$ distribution through $q$\\
        \hline
        $\mu_{m}$ & $\mathcal{U}(20 \text{M}_\odot, 100 \text{M}_\odot)$ & 
        Mean of Gaussian peak in $m_1$ distribution\\
        \hline
        $\sigma_{m}$ & $\mathcal{U}(1 \text{M}_\odot, 10 \text{M}_\odot)$ & 
        Standard deviation of Gaussian peak in $m_1$ distribution\\
        \hline
        $f_{p}$ & $\mathcal{U}(0, 1)$ & 
        Fraction of $m_1$ in the Gaussian peak\\
        \hline
        $\lambda$ & $\mathcal{U}(-5, 4)$ & 
        Index for power-law component of $m_1$ distribution\\
        \hline
        $\gamma$ & $\mathcal{U}(-2, 10)$ & 
        Index for power-law distribution of $q$\\
        \hline
        $m_{\max}$ & $\mathcal{U}(60 \text{M}_\odot, 100 \text{M}_\odot)$ & 
        Maximum possible BH mass\\
        \hline
        $m_{\min}$ & $\mathcal{U}(2 \text{M}_\odot, 10 \text{M}_\odot)$ & 
        Minimum possible BH mass\\
        \hline
        $\kappa$ & $\mathcal{U}(-100, 100)$ & 
        Determines the level of correlation between $q$ and $\chi_\text{eff}$\\
        \hline
    \end{tabular}
    \caption{List of hyper-parameters used in the model defined in Section~\ref{sec:model_and_implementation} along with their respective priors. Here, $\mathcal{U}(a,b)$ indicates a uniform distribution on the interval $[a,b]$.
    }
    \label{tab:priors}
\end{table*}

In this section, we construct a copula model for $\pi(q, \chieff)$.
The first step is to choose marginal distributions for $\pi(q)$ and $\pi(\chieff)$, which we choose to match the the correlated model from \cite{Callister_2021}.
The distribution of mass ratio is therefore taken to be a power-law; see equation~\ref{eq:p_q_m1}.
There is some complication because the distribution of $q$ is subtly conditioned on primary mass $m_1$---since we require each component mass to be on the interval $(m_\text{min}, m_\text{max})$, not all values of $q$ are allowed for some values of $m_1$.
We therefore marginalize numerically over $m_1$ \footnote{Primary mass is distributed according to a power-law and Gaussian peak mixture model such that $\pi(m_1|\Lambda)$ follows equation~\ref{eq:p_m1}.}.
The distribution of $\chieff$ is obtained numerically by marginalizing over $q$ in equation~\ref{eq:gaussian_p_chi_eff}; see the right-hand panel of Fig.~\ref{fig:callister+}.
Using the marginal distributions of $(q, \chieff)$, we compute the copula variables $(u,v)$ following Eqs.~\ref{eq:example_u}-\ref{eq:example_v}.

The next step is to choose a copula.
We opt to use the Frank copula density function:
\begin{equation}
\label{eq:frank_copula}
    \pi_c(u,v|\kappa) = \frac{\kappa e^{\kappa(u+v)}(e^\kappa - 1)}{\left(e^\kappa - e^{\kappa u} - e^{\kappa v} + e^{\kappa(u + v)}\right)^2} ,
\end{equation}
which we chose by trial and error on the grounds that it produces physically plausible-looking correlations (to our eyes) in the $(q, \chieff)$ plane; see the right four panels of Fig.~\ref{fig:copula_examples}.
The subjectivity associated with the choice of copula is similar to other model-building choices, e.g., the functional form of $\pi(\chieff|q)$.
However, in principle, one may perform model selection to choose between copulas.
Finally, we use equation~\ref{eq:x-y_to_u-v} to define $\pi_c(q, \chieff | \Lambda, \kappa)$.

By construction, our model is similar to the model from \cite{Callister_2021}.
The two models share hyper-parameters, though, our model has an additional parameter $\kappa$.
There is, however, an important difference.
In \cite{Callister_2021}, the $\alpha$ parameter controls the correlation between $(q, \chieff)$---while also inadvertently changing the marginal distribution for $\chieff$.
However, in our model, $\alpha$ is used only to change the marginal distribution for $\chieff$.
The same can be said of $\beta$.
The covariance between $(q, \chieff)$ is determined entirely with $\kappa$.

With our model now defined, the next step is to carry out population inference to obtain posterior distributions for the hyper-parameters---especially $\kappa$.
With each sampling step, the copula variables $(u,v)$ are reevaluated to reflect the newly proposed values $\Lambda$, and the corresponding marginal distributions $\pi(q|\Lambda)$ and $\pi(\chieff|\Lambda)$.
We employ the population inference package \texttt{GWPopulation} \citep{Talbot_2019}, which employs \texttt{Bilby} \citep{bilby,bilby_gwtc1}.
We use the nested sampler \texttt{DYNESTY} \citep{Speagle_2020}.
We use the same 44 BBH events from GWTC-2 \citep{GWTC2, data} as \cite{Callister_2021}.
In reweighting this data, we utilise the fiducial prior for $\chieff$ derived in \cite{Callister_2021b}.
The priors for $\Lambda$ are listed in Table~\ref{tab:priors}.

\section{Results}
\label{sec:results}

\begin{figure*}
    \centering
    \includegraphics[width=\textwidth]{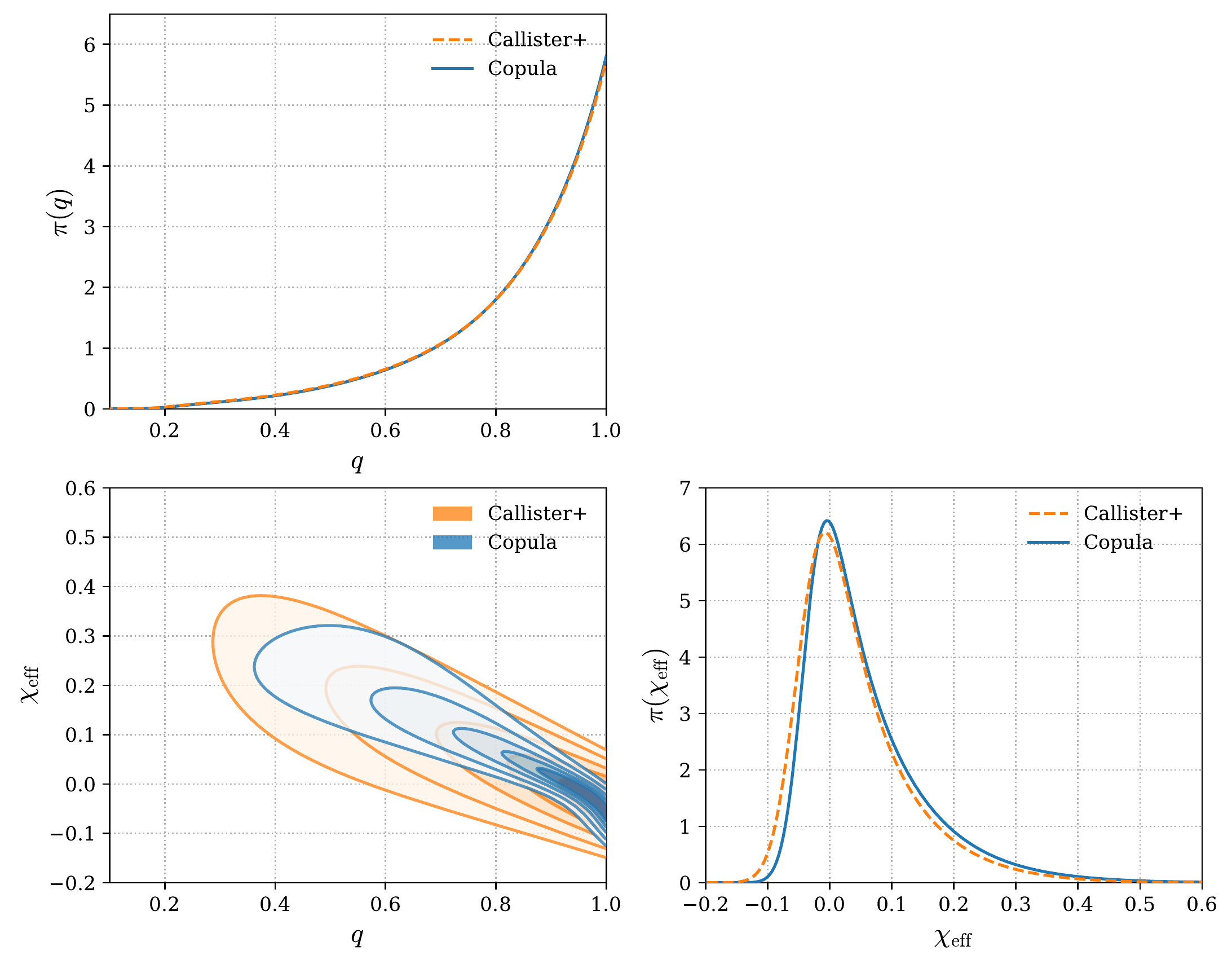}
    \caption{Reconstructed distributions for the Callister \textit{et al.} (2021) model (orange) and the copula model from Section~\ref{sec:model_and_implementation} (blue) using maximum posterior values for all hyper-parameters. \textbf{Top left}: the marginal distributions of mass ratio. \textbf{Bottom right}: the marginal distributions of effective inspiral spin. \textbf{Bottom left}: the joint distributions for mass ratio and effective inspiral spin.}
    \label{fig:compare_ppds}
\end{figure*}

\begin{figure*}
    \centering
    \includegraphics[width=\textwidth]{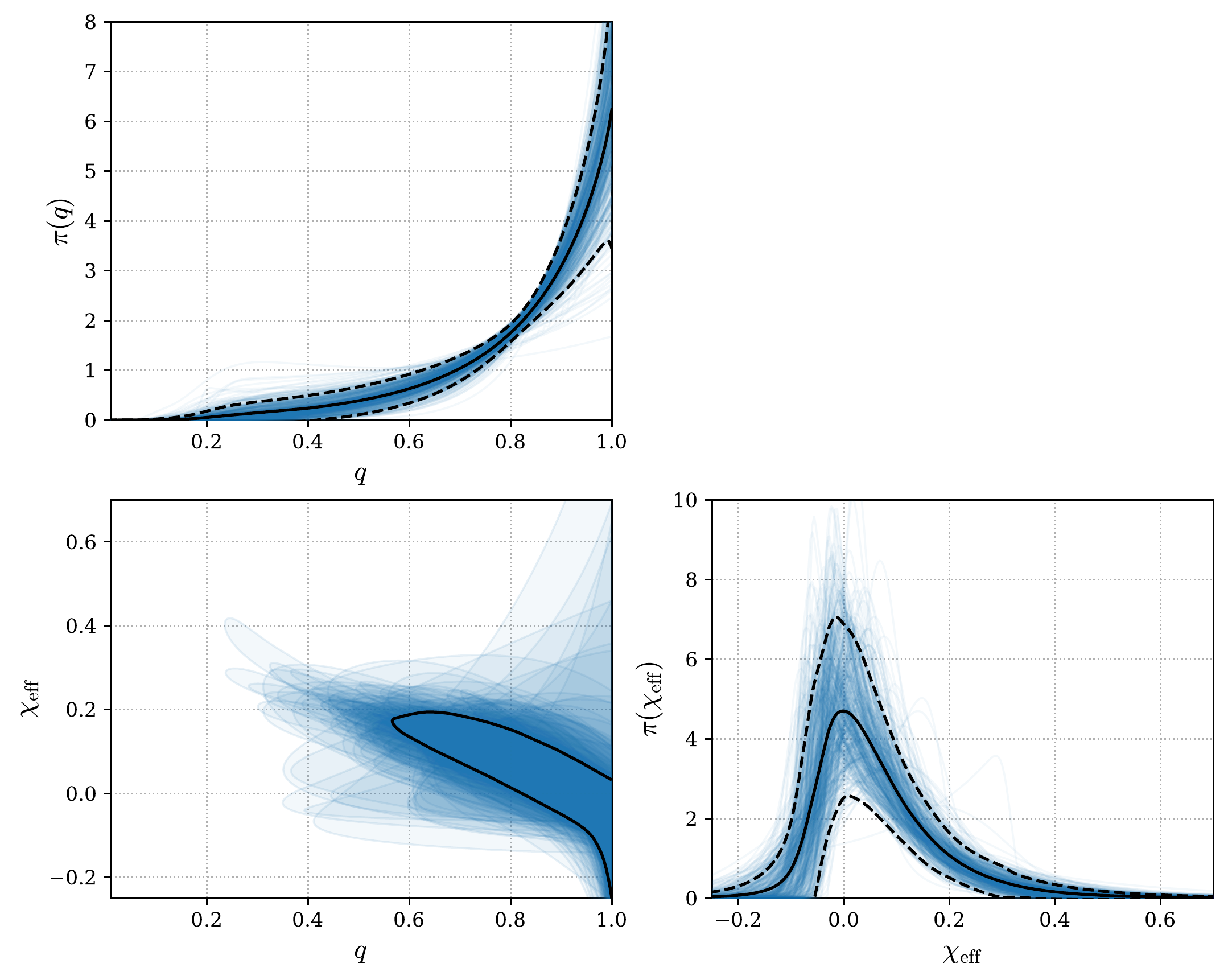}
    \caption{Posterior predictive distributions for the copula model presented in Section~\ref{sec:model_and_implementation}. \textbf{Top left}: the marginal distribution of mass ratio. \textbf{Bottom right}: the marginal distribution of effective inspiral spin. \textbf{Bottom left}: the joint distribution for mass ratio and effective inspiral spin. In the one-dimensional plots, the blue lines trace distributions implied by 500 random draws from the posteriors of the hyper-parameters. The solid black lines shows the mean distribution, whilst the dashed lines encapsulate the 90\%-credible regions. In the two-dimensional plot, the blue segments show 90\% probability regions for the population implied by individual hyper-parameter draws. The mean of these regions is outlined in black.}
    \label{fig:sampled_ppd}
\end{figure*}

\begin{figure*}
    \centering
    \includegraphics[width=\textwidth]{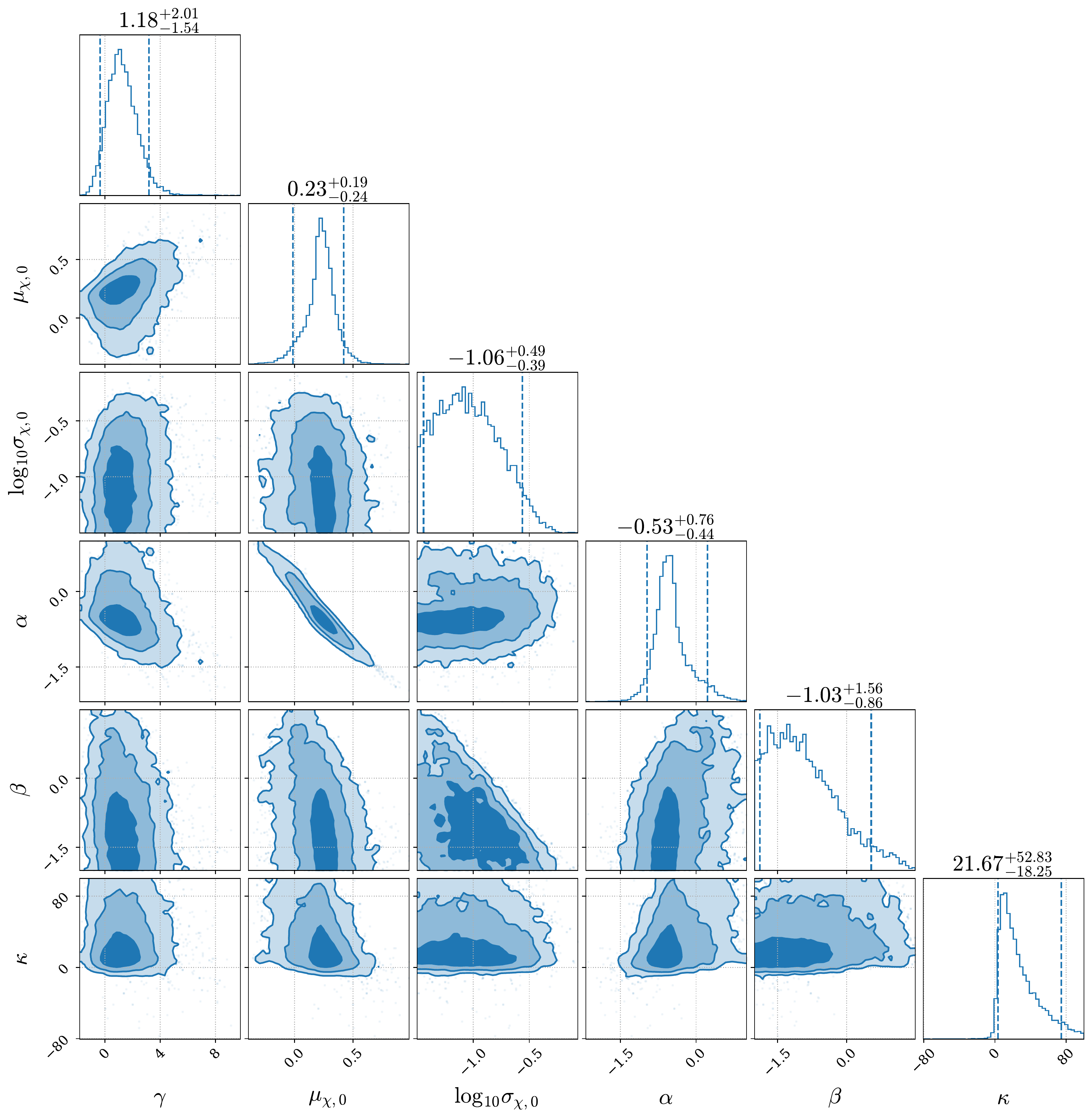}
    \caption{Corner plot of the posteriors for hyper-parameters governing the distribution of mass ratio and effective inspiral spin. Contours on the two-dimensional plots, from darkest to lightest, show 50, 90 and 99\%-credible regions, whilst blue dashed lines on one-dimensional plots show 90\%-credible regions. Values listed above give the median of the posterior as well as the aforementioned 90\%-credible region.}
    \label{fig:corner}
\end{figure*}

In Fig.~\ref{fig:compare_ppds}, we show in blue the reconstructed marginal distributions of $q$ and $\chi_\text{eff}$, as well as the associated joint distribution obtained from our fit.
Overlayed in orange is the maximum-posterior fit from \cite{Callister_2021}.
As expected, the marginal distribution of mass ratio is effectively identical in either case.
We recover a similar marginal distribution for $\chieff$ when modelling its shape separately from the correlation.
In hindsight, this is perhaps unsurprising given the preference for this asymmetric shape seen in previous work \citep{Roulet_2021, Galaudage_2021b}.
Turning our attention to the joint distribution, we find support for an anti-correlation between mass ratio and effective inspiral spin similar to the one inferred by \cite{Callister_2021}.
This indicates that the anti-correlation described by \cite{Callister_2021} is unlikely to be just a byproduct of the marginal distribution $\pi(\chieff)$ \footnote{
The joint distribution implied by the copula model also exhibits a subtle widening as q approaches unity.
As can be seen in Fig.~\ref{fig:copula_examples}, copulas tend to exhibit regions of relatively high density in the corners in order to keep the marginal distributions of $u$ and $v$ flat near 0 and 1.
This can lead to kinks in joint distributions constructed with copula density functions, such as the one seen here.
}.
In Fig.~\ref{fig:sampled_ppd}, we show the same reconstruction for the copula model, this time using the mean constructed via 500 hyper-parameter draws from their associated posteriors.

Whilst the anti-correlation hypothesis is well supported, we cannot entirely rule out the possibility of no correlation at high credibility.
Indeed, examining the posterior distribution for $\kappa$ in Fig.~\ref{fig:corner}, we find that $\kappa = 0$ is ruled out with 98.7\% credibility, the same significance obtained in \cite{Callister_2021} \footnote{
Note that the support for $\kappa = 100$ (the upper bound of the prior) is non-zero.
This raises the question of whether it is necessary to extend the bounds of the prior on $\kappa$.
However, the correlation does not change much beyond values of $\kappa = 100$, which is already an extremely tight anti-correlation.
Extending the prior boundary results in a long, low-density tail.
This yields a subtle $\sim 0.1\%$ increase in the credibility for anti-correlation.
}.
Meanwhile, we obtain credible intervals of $\alpha = -0.53^{+0.76}_{-0.44}$ and $\beta = -1.03^{+1.56}_{-0.86}$, which are similar to the values found in \cite{Callister_2021}: $\alpha = -0.46_{-0.28}^{+0.29}$ and $\beta = -0.83_{-1.01}^{+1.28}$.
This is in spite of the fact that, in our model, $\alpha$ and $\beta$ have no effect on the $(q, \chieff)$ covariance---they only affect the shape of the $\chieff$ distribution.

\section{Discussion}
\label{sec:discussion}
We find some evidence in support of the idea that mass ratio and effective inspiral spin are anti-correlated, corroborating the findings of \cite{Callister_2021}.
Our results demonstrate that the significance of \cite{Callister_2021}'s anti-correlation is not arising entirely from an improvement in the $\chieff$ fit.
This has potential implications for the formation channels of BBH systems observed in gravitational waves.
Several of these are already discussed in \cite{Callister_2021}, but we discuss them here for completeness.

First, simulations by \cite{Broekgaarden_2022} suggest that BBH systems, which undergo mass ratio reversal (where the second born BH accretes enough mass to become the heavier BH in the binary) may exhibit anti-correlation in the $q - \chieff$ plane.
Similar anti-correlation is predicted by \cite{Bavera_2021} when assuming BBH systems are formed through either common envelope evolution with high common envelope efficiency, or through stable mass transfer using super-Eddington accretion.
The preference for a qualitatively similar anti-correlation seen here may indicate that a large portion of BBH merger events observed to date come from one or more of the three aforementioned scenarios.

Another potential explanation for this feature, proposed by \cite{Callister_2021}, is that the anti-correlation is produced via the Simpson's paradox \citep[see][]{Blyth_1972}.
According to this hypothesis, the population of BBH merger events does not exhibit anti-correlation locally, but it appears globally due to groupings of separate sub-populations in the $q$-$\chi_\text{eff}$ plane, potentially born through a mixture of different formation channels.
Through model selection, it may be possible to distinguish between these two scenarios.

In a follow up to \cite{Callister_2021}, \cite{Mckernan_2022} suggest that a $(q,\chieff)$ anti-correlation may arise from BBH systems forming dynamically in AGN, provided several assumptions hold: black holes in the AGN disk are heavier and have aligned spins whilst those outside are lighter and have random spin alignments; the inner AGN disk is dense but small; and migration of black holes into the disk is turbulent.
If one could show this phenomenological anti-correlation aligns with these predictions, the above properties may apply to AGN more generally \citep{agn_avi}.

Given that the definition of $\chieff$ is dependent on $q$ (see equation \ref{eq:chi_eff}), one might expect some degree of covariance in $(q, \chieff)$ is implied even if component spin parameters are distributed independently.
We investigate this in Appendix Section \ref{sec:appendix}, finding that the level of correlation exhibited in our results cannot be implied by the parameterisation of $\chieff$ alone.

We now turn our attention from the astrophysical implications of our results to the use of copulas in gravitational-wave astronomy.
There are many different existing copula density functions.
This provides the potential to probe not only the level of correlation between two parameters $x$ and $y$, but also the precise shape of the correlation in the $x-y$ plane.
Model comparisons for marginal distributions such as those in \cite{GWTC1_analysis} \citep[see also the tutorial by][]{Thrane_2019} can be simply adapted to compare the fit of several different copula density functions.
This can be used to look for subtle features in the covariance.

It is natural to ask if the principles described in this paper can be extended to look for correlations in $>2$ dimensions.
Progress has been made creating three-dimensional copulas \citep[e.g.][]{Sklar_1996, Bedford_2002, Aas_2009, Devroye_2010}, although the vast majority of literature on this topic involves correlations in just two dimensions.
Nevertheless, it is possible to identify complicated correlations in large-dimension spaces using a formalism like this.
One method for doing so is through use of `vine copulas' \citep{Bedford_2002}.
This framework effectively stitches together multiple two-dimensional copulas, allowing for an infinitely scalable n-dimensional copula \citep[see][for an example applied to galactic mass-luminosity functions]{Takeuchi_2020}.

We see copulas as essential for model testing, but not necessarily for model building.
The `dependent-distribution' models used in \cite{Callister_2021}, \cite{Safarzadeh_2020}, and \cite{Biscoveanu_2022} are all, in our view, physically reasonable.
However, having identified that a correlation might be present with such a dependent-distribution model, follow-up study with a copula is indispensable in order to determine if the correlation is bona fide, or due to a improvement in the dependent-parameter marginal distribution.

Given the large number of questions and competing theories for BBH formation, the study of covariance between astrophysical parameters is likely to be useful for future investigations.
Copulas are effective tools for measuring covariance in gravitational-wave astronomy.
They allow us to eliminate extraneous and potentially confounding changes to models when testing for correlation, with results that can be easily interpreted via a single parameter.
Equally appealing is their ease of use, effectively acting as modular additions to existing models.

\section*{Acknowledgements}
We thank Tom Callister, Maya Fishbach, and our anonymous referee for insightful comments and discussion of the work presented here.
We also thank Shanika Galaudage for her assistance in utilising the \texttt{GWPopulation} package.
We acknowledge support from the Australian Research Council (ARC) Centre of Excellence CE170100004.
This material is based upon work supported by NSF's LIGO Laboratory which is a major facility fully funded by the National Science Foundation.
The authors are grateful for computational resources provided by the LIGO Laboratory and supported by National Science Foundation Grants PHY-0757058 and PHY-0823459.
This research has made use of data or software obtained from the Gravitational Wave Open Science Center (gw-openscience.org), a service of LIGO Laboratory, the LIGO Scientific Collaboration, the Virgo Collaboration, and KAGRA. LIGO Laboratory and Advanced LIGO are funded by the United States National Science Foundation (NSF) as well as the Science and Technology Facilities Council (STFC) of the United Kingdom, the Max-Planck-Society (MPS), and the State of Niedersachsen/Germany for support of the construction of Advanced LIGO and construction and operation of the GEO600 detector. Additional support for Advanced LIGO was provided by the Australian Research Council. Virgo is funded, through the European Gravitational Observatory (EGO), by the French Centre National de Recherche Scientifique (CNRS), the Italian Istituto Nazionale di Fisica Nucleare (INFN) and the Dutch Nikhef, with contributions by institutions from Belgium, Germany, Greece, Hungary, Ireland, Japan, Monaco, Poland, Portugal, Spain. The construction and operation of KAGRA are funded by Ministry of Education, Culture, Sports, Science and Technology (MEXT), and Japan Society for the Promotion of Science (JSPS), National Research Foundation (NRF) and Ministry of Science and ICT (MSIT) in Korea, Academia Sinica (AS) and the Ministry of Science and Technology (MoST) in Taiwan.

\section*{Data Availability}
We analyse data from \cite{GWTC2}, made publicly available through \cite{data}.
This work also presents results from \cite{Callister_2021}.
The modified version of \texttt{GWPopulation} used in this analysis is publicly available at \url{https://github.com/ChristianAdamcewicz/gwpopulation/tree/q_chi_eff_cal_copula}.

\bibliographystyle{mnras}
\bibliography{refs}

\appendix

\section{Implied correlations from component spin models}
\label{sec:appendix}

It is not  uncommon to model the distributions of binary black hole component spin magnitudes $\chi_{1,2}$ and tilts $t_{1,2}$ instead of the $\chieff$ distribution \citep[see, for example][]{GWTC2_analysis,GWTC3_analysis,Talbot_2017,Wysocki_2019,Galaudage_2021b}.
Given that the definition of $\chieff$ depends on the mass ratio $q$ (see equation \ref{eq:chi_eff}), one might wonder if models described in terms of independently distributed physical parameters,
\begin{align}
    \pi(q,\chi_1,\chi_2, ...) = \pi(q)\pi(\chi_1)\pi(\chi_2) ...
\end{align}
imply (inadvertent) correlation between $(q, \chieff)$.
We explore this possibility using the \textsc{Default} model described in \cite{GWTC3_analysis} \citep[see also][]{Talbot_2017, Wysocki_2019}.
Using distributions reconstructed with the best-fit hyper-parameters, we draw $2.5 \times 10^{4}$ samples for $(q, \chi_1, \chi_2, \cos t_1, \cos t_2)$, and then convert these to $\chieff$ samples using equation \ref{eq:chi_eff}.

\begin{figure*}
    \centering
    \includegraphics[width=\textwidth]{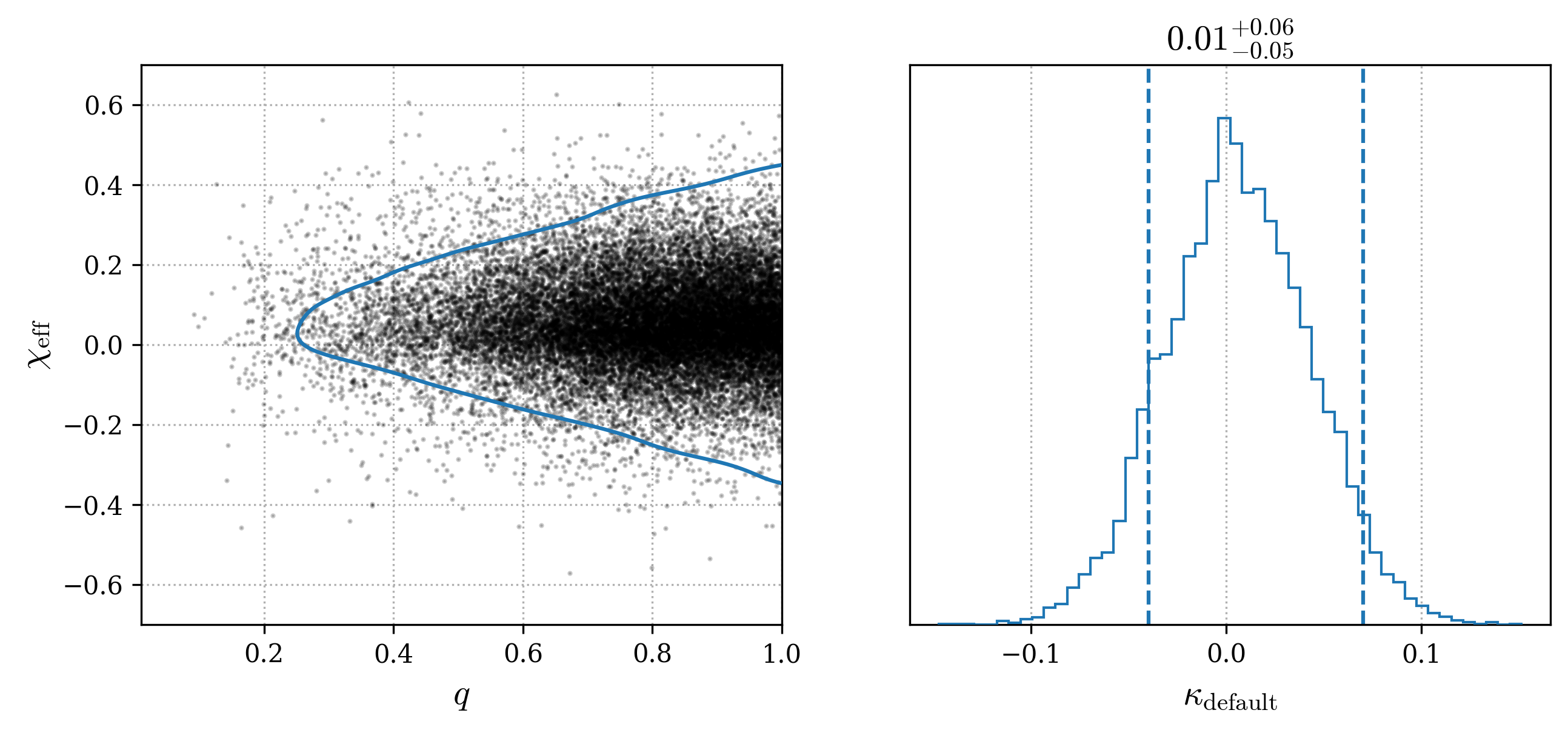}
    \caption{Correlation between mass ratio and effective inspiral spin implied by the independently distributed component spins of the \textsc{Default} model. \textbf{Left}: joint distribution of mass ratio and effective inspiral spin implied by the \textsc{Default} model. Samples generated using the model are shown in black. The blue line shows the reconstructed distribution using the best-fit value for the correlation parameter $\kappa_\text{default}$. \textbf{Right}: posterior distribution for $\kappa_\text{default}$. The blue dashed lines show the 90\% credible regions. These along with the median of the posterior distribution are listed above.}
    \label{fig:default_q_chi_eff}
\end{figure*}

We then measure the correlation between $q$ and $\chieff$ in these samples.
To do so, we make use of the known marginal distribution for $q$, and a suitable approximation of the marginal distribution for $\chieff$ obtained using a Gaussian kernel density estimate.
We then construct a simple model for the joint distribution of $(q, \chieff)$ by linking the known marginal distributions with a Frank copula density function dependent on a correlation parameter $\kappa_\text{default}$ (see Section \ref{sec:an_introduction_to_copulas} for details).
This allows us to perform a simple one-dimensional fit for $\kappa_\text{default}$ -- the correlation in the $q - \chieff$ plane implied by the \textsc{Default} model.
The results of this analysis are shown in Fig.~\ref{fig:default_q_chi_eff}.

We find that the \textsc{Default} model does not imply a strong correlation between $q$ and $\chieff$.
This suggests that the $(q, \chieff)$ correlation observed by \cite{Callister_2021}, and discussed in Section~\ref{sec:results}, is not a predetermined consequence of the parameterisation of $\chieff$.

\bsp
\label{lastpage}
\end{document}